\documentclass{mem}
\usepackage{natbib}\usepackage{txfonts}\usepackage{balance}
\usepackage{flushend}
\usepackage{graphicx}
\usepackage[breaklinks,pdftex]{hyperref}
\idline{75}{282}
\begin{document}

\title{
Insights into Extragalactic Background Light constraints with MAGIC archival data
}


\author{
R. \,Grau\inst{1} 
\and A. \, Moralejo\inst{1} for the MAGIC collaboration.
          }

\institute{
Institut de Física d’Altes Energies (IFAE), The Barcelona Institute of Science and Technology (BIST), E-08193 Bellaterra (Barcelona), Spain  \and \\
\email{rgrau@ifae.es}\\
}

\authorrunning{Grau}

\titlerunning{Extragalactic Background Light}

\date{Received: XX-XX-XXXX (Day-Month-Year); Accepted: XX-XX-XXXX (Day-Month-Year)}

\abstract{

The Extragalactic Background Light (EBL) is the accumulated light emitted throughout the history of the universe, spanning the UV, optical, and IR spectral ranges.
Stars and dust in galaxies are expected to be the main source of the EBL. However, recent direct measurements performed beyond Pluto's orbit (less affected by foregrounds than those performed from the Earth) hint at an EBL level in the optical band larger than the one expected from the integrated contribution of known galaxy populations.
One approach that could solve this controversy uses Very High Energy (VHE) photons coming from sources at cosmological distances. These photons can interact with the EBL producing electron-positron pairs, a process that leaves an imprint on the observed VHE spectrum. This technique, however, requires assumptions on the intrinsic spectrum of the source, which can compromise the robustness of EBL constraints.
In this contribution, we used Monte Carlo simulations and archival data of the MAGIC telescopes to study the impact that the assumptions adopted in the literature have on the estimate of the EBL density, and how using more generic ones would modify the results.
Our results show how the EBL density constraints obtained highly depend on the intrinsic spectral shape assumed for the source. We have studied two different methods to reduce the assumptions on the intrinsic spectral shape to get more robust results. This will be especially important for upcoming with new VHE facilities, where systematic uncertainties are expected to play a more significant role compared to statistical ones.

}
\maketitle{}

\section{Introduction}
The Extragalactic Background Light (EBL) is all the light produced since the beginning of the universe, covering a range of wavelengths from ultraviolet (UV) to optical and infrared (IR). Most of this light is produced by stars, either as direct emission (cosmic optical background, COB) or re-radiated after being absorbed by dust (cosmic infrared background, CIB) \citep{Cooray_2016}. \\
Different methods exist for probing the EBL. Direct measurements are difficult since it is much fainter than other local foregrounds at the same wavelength, such as zodiacal light. One possibility to reduce those foregrounds is to take data far away from the Sun. This is what recent studies \citep{Lauer_2022} have done, where they used data from the space telescope LORRI, onboard of the New Horizons spacecraft and beyond Pluto's orbit, to reduce these foregrounds. They isolated the COB at a pivot wavelength of $0.608$ $\mu m$ with a result larger (and in tension with) than results obtained with other methods. However, a reanalysis of the data \citep{Postman2024} showed a lower result, compatible with galaxies being the only source of COB. Another method uses the estimated number of galaxies, obtained from deep surveys, with galaxy formation and evolution models to estimate the integrated flux. The downside of this method is that it is not sensitive to truly diffuse or unknown components of EBL. The method that we are revising is the gamma-ray-based one. For this method, we use the fact that Very High Energy (VHE) gamma rays coming from sources at cosmological distances interact with EBL, producing an electron-positron pair. This interaction can be observed in the source's VHE spectra as an energy-dependent absorption imprint. We can use this to probe EBL, independently of its origin, but making assumptions on the source's intrinsic gamma-ray spectra. The intrinsic spectra of the source represents how we would see the source from Earth if there was no EBL (the redshifted gamma-ray spectra of the source).

Previous results like the one from \cite{Acciari2019}, use simple concave functions to fit the spectra of the sources and do a profile likelihood of the EBL density scale factor ($\alpha$) about a certain EBL model, following the formula:

\begin{equation}
    \frac{dF_ {obs}}{dE} = \frac{dF_{int}}{dE} e^{-\alpha\tau(E,z) }
\end{equation}
where $\alpha$ accounts for EBL absorption, E is the energy of arrival at Earth (red-shifted), $\frac{dF_{int}}{dE}$ is the source's red-shifted intrinsic spectrum, and $\tau(E,z)$ is the model's EBL optical depth. 
This works typically use Wilks' theorem \citep{Wilks_1938} to compute the $\alpha$ constraint from the profile likelihood of $\alpha$. However, it is not guaranteed that the conditions to apply this theorem are fulfilled. For example, the P-values of the best fits are often very low, hinting the possible presence of hidden systematic errors. Also parameters reaching limits (like the concavity limit) could be a problem for Wilks' theorem.
Another problem with this method could be the selection of the fit function, since the results obtained depend on this selection.

Our goal is to make a Monte Carlo simulation to check the coverage of the results, checking the validity of Wilks' theorem. The same simulation can be used to compute more accurate uncertainties in case the coverage differs significantly from the expected $68\%$.
Additionally, we will try to use weaker assumptions on the intrinsic shape of the source.

For that purpose, we will use Mrk421 and 1ES1011+496 MAGIC archival data. Mrk421 and 1ES1011+496 are 2 BLLac objects at $z=0.03$ and $z=0.212$ respectively, which emit VHE gamma rays, making them good sources for studying EBL. The MAGIC  telescopes (\cite{Aleksi__2016}) are a system of two IACTs at the Roque de los Muchachos Observatory on La Palma, Spain. These telescopes are designed to detect very high-energy (VHE) gamma rays in the range of 30 GeV to 100 TeV by observing the Cherenkov radiation produced when gamma rays interact with Earth's atmosphere.
\section{Toy Monte-Carlo simulation}
In order to test the validity of Wilks' theorem and compute more accurate uncertainties we have made a Monte Carlo simulation of MAGIC observations. It can simulate multiple Poisson realizations of the same spectra modeled by a function.
Every realization is then analyzed using a Poissonian likelihood maximization, in the same way real data is analyzed. The likelihood has one term for each bin in reconstructed energy (i), of the form:

\begin{equation}
\begin{array}{l}
    L_{i}(ebl, \theta) = Poisson(g'_{i}(ebl, \theta) +b_{i}; N_{on,i})\,\cdot \\
    Poisson(b_{i}/\beta; N_{off,i})\cdot Gauss(g'_{i};g_{i},\Delta g_{i})
\end{array}
\label{eq:likelihood}
\end{equation}
where $N_{on, i}$ and $N_{off, i}$ denote the recorded events in reconstructed energy bins (i = 1, ..., $N_{bins}$). $N_{on}$ corresponds to events around the source, while $N_{off}$ represents background events in three control regions. $g_i$ is the Poisson parameter (mean number) of gammas in the ON-source region, $b_{i}$ is the Poisson parameter of the background in the ON-source region and is treated as a nuisance parameter. The factor $\beta$ is the ratio of ON to OFF exposure which in this case is $\beta = 1/3$. $g_{i}'$ is a nuisance parameter accounting for uncertainty in the instrument response function, while $\Delta g_i$ represents the range of values associated with that uncertainty.\footnote{For more details on the likelihood see Appendix A in \cite{Acciari2019}}
From this, we obtain multiple profile likelihoods for the $\alpha$ parameter, one from each realization, from $\alpha = 0$ to $2$ in steps of $0.05$. For each realization we check the difference of -2logLikelihood (-2logL) between the minimum -2logL and the true (simulated) value of $\alpha$ ($\alpha = 1$). We can calculate the cumulative distribution function (CDF) of this difference, and, if Wilks' theorem is applicable it should follow a $\chi^2$ with 1 degree of freedom.

Comparing the P-values of the simulation with the ones obtained with real data, we saw that the real data ones were much smaller. This could hint the presence of hidden systematics. We decided to model these systematics in the simulation as Gaussian systematics in the effective area of the telescope, independent in each energy bin but with the same standard deviation (in relative terms).

\begin{figure}[h!]
\vspace{-0.91cm}
    \centering
    \includegraphics[width=\linewidth]{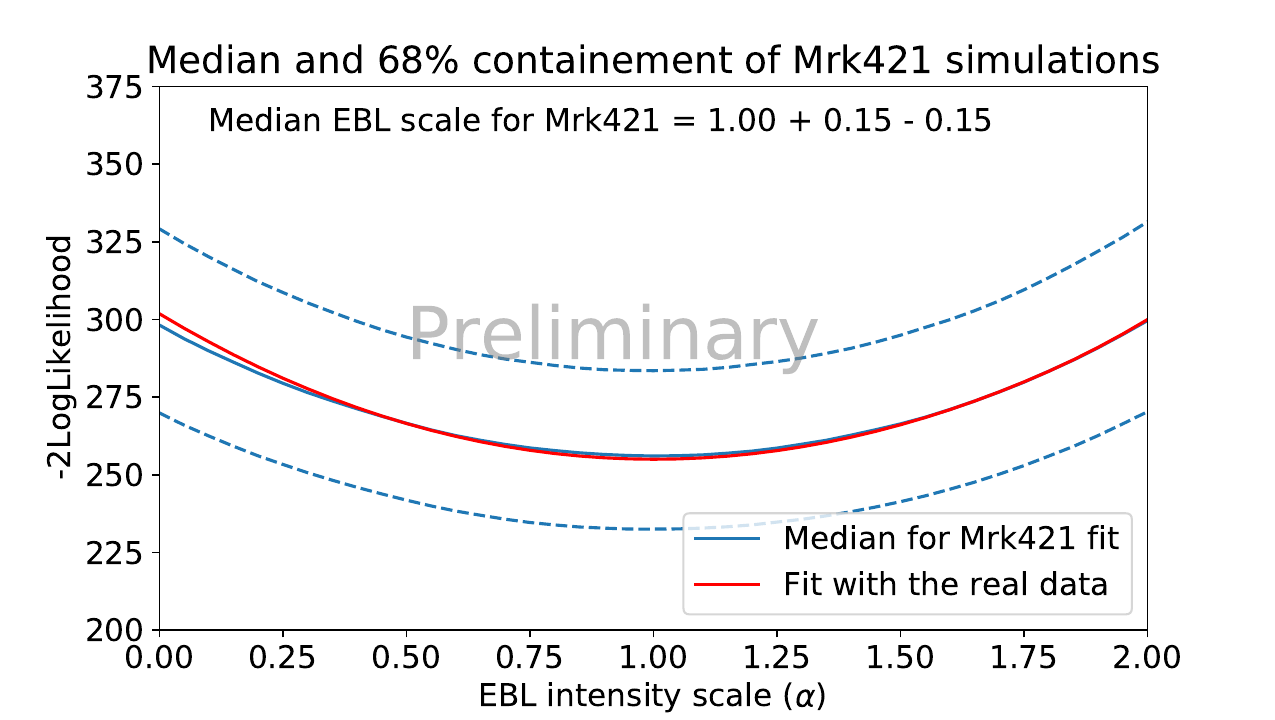}
    \caption{Results from $10^4$ realizations of the simulation for the 15 Mrk421 spectra with the D11 model. In red the results of the real data fits.}
    \label{fig:Mrk_sim}
\vspace{-12pt}
\end{figure}

We used the Monte Carlo simulation to generate 10k observations emulating the 15 Mrk421 spectra from \cite{Acciari2019}. We used the best-fit function of the paper as the intrinsic spectra of each dataset and the EBL model from \cite{Dominguez_2011} (D11 in the following). The result of computing the combined profile likelihood of all the realizations is shown in \autoref{fig:Mrk_sim}.

\begin{figure}[h]
\vspace{-0.7cm}
    \centering
    \includegraphics[width=\linewidth]{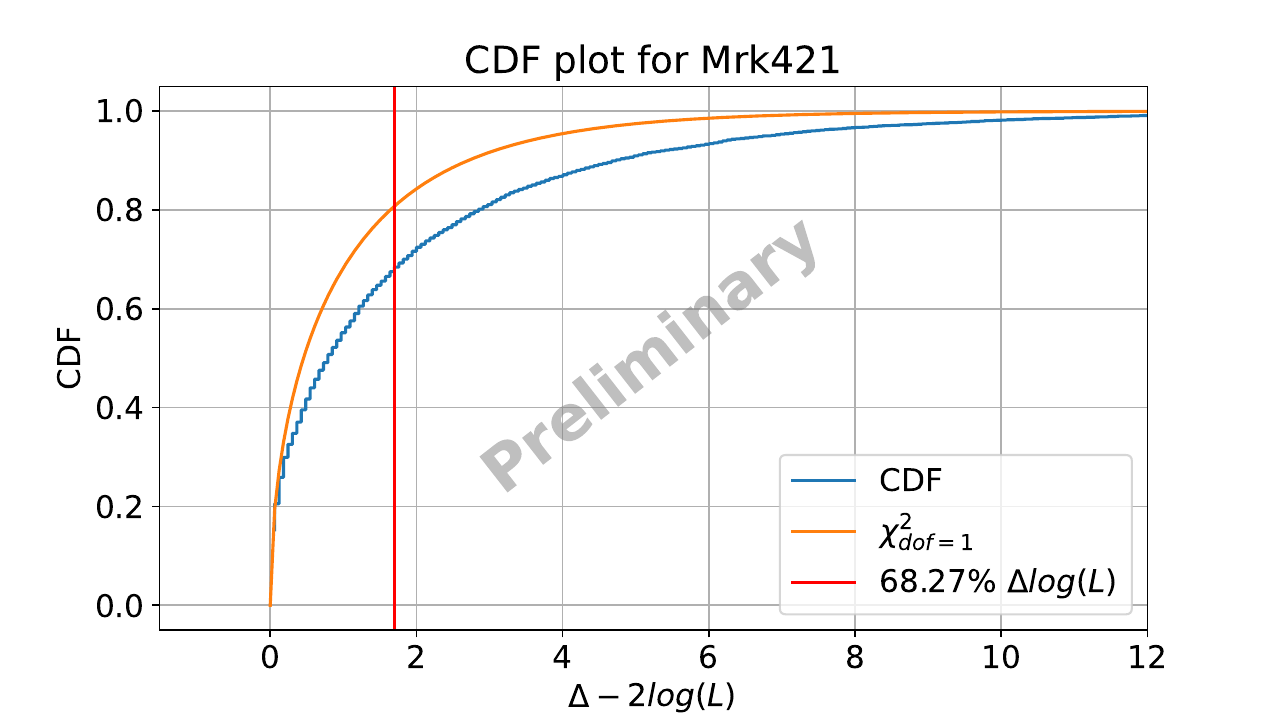}
    \caption{Cumulative distribution function of the simulation compared to a $\chi^2$ distribution. The vertical red line shows the point where the CDF equals $68.27\%$}
    \label{fig:Mrk_CDF}
\end{figure}

With this profile likelihoods, we can now plot the CDF of the difference of -2logL between the minimum of each realization and its value at the true value of $\alpha$ $(\alpha=1)$ (shown in \autoref{fig:Mrk_CDF}). We can see how the CDF differs from the $\chi^2$ with 1 degree of freedom, which implies that Wilks' theorem should not be used here to compute uncertainties and therefore the $\Delta$(-2logL)$=1$ to get the $68\%$ Confidence Level (CL) of the results (the real data ones). Instead, we should use the $\Delta$-2logL where the CDF equals to $0.68$. In this case, the $\Delta$-2logL required is $1.70$.

\begin{figure}
    \centering
    \includegraphics[width=\linewidth]{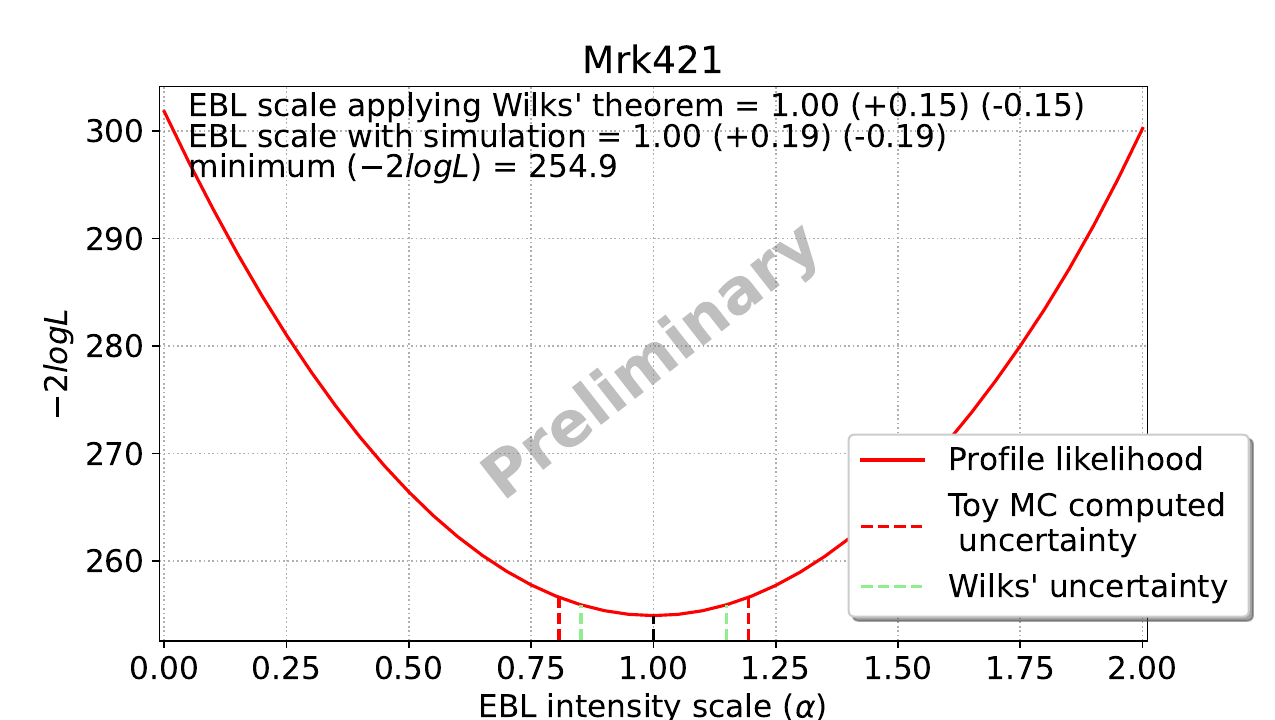}
    \caption{Combined profile likelihood of the 15 Mrk421 real data spectra. Uncertainties computed using Wilks' theorem are shown in light green and uncertainties obtained using our simulations are shown in red.}
    \label{fig:Mrk421_real}
\end{figure}

Using this result, we made the profile likelihood of the real data and compared the uncertainties estimated using Wilks' theorem with those derived from our simulations (\autoref{fig:Mrk421_real}). The comparison revealed a $27\%$ increase in uncertainty, indicating that previous studies had underestimated the uncertainties.
This shows the importance of accounting for systematic effects in future analyses to ensure robust and accurate uncertainty estimates.

\section{New methods}
BLLac objects are not expected to have inflection points in their VHE spectra (after the Inverse Compton peak). The EBL transmissivity ($e^{-\tau}$), on the other hand, has a \textit{wiggle} at about $1$ TeV. That's why, to reduce the number of assumptions on the intrinsic spectra of the source, we have developed two methods to determine the EBL by looking for this \textit{wiggle}.

\subsection{Generic concave function}
The first method uses a generic concave function that can, in principle, fit all the curvature of the intrinsic spectrum and the EBL transmissivity but the inflection point. The function we have chosen for this purpose is the Multiply Broken Power-Law (MBPWL), which is a Power-Law (PWL) defined by parts, where the photon index increases in points called nodes or knots (to impose concavity). To make computation easier and avoid non-convergences, we select the number of nodes, the position of the first and last node, and the rest of the nodes are logarithmically spaced between the first and last one. 

\begin{figure}
    \centering
    \includegraphics[width=\linewidth]{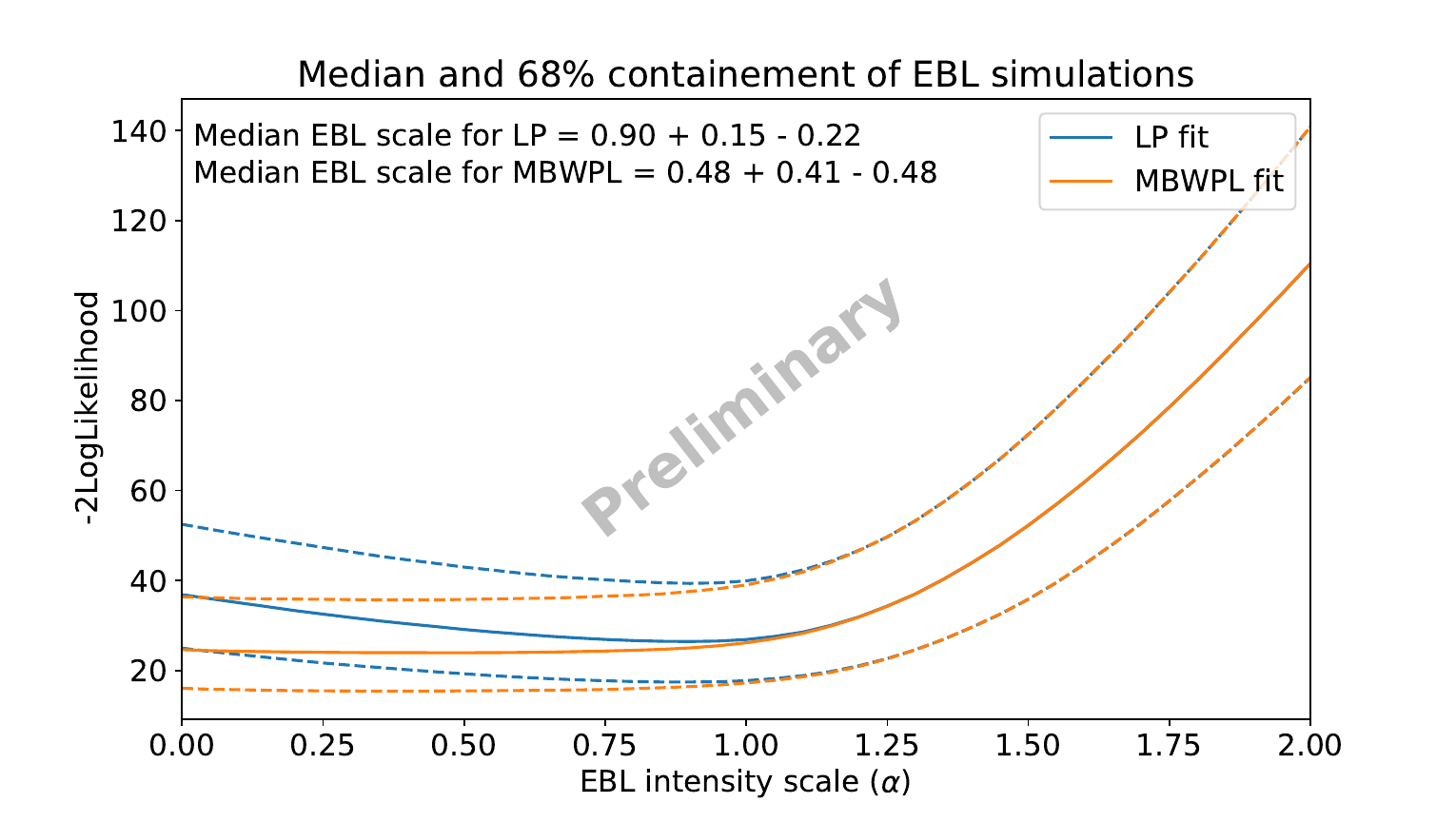}
    \caption{Median of the profile likelihood results for 1ES1011+496 simulated by a PWL and fitted with a LP and a MBPWL using D11 EBL model. In dotted lines the region containing 68\% of the realizations around the median.}
    \label{fig:1ES_MBPWL_prof}
    \vspace{-0.4cm}
\end{figure}

We have simulated a spectrum like the one of 1ES1011+496\footnote{A PWL with photon index $\Gamma = 2.03$ and normalization factor at 250 GeV $F_0 = 8.70 x 10^{-6} m^{-2} s^{-1} TeV^{-1}$, with the source located at $z=0.212$ and observed during $11.8h$.} and have run 10k realizations of it. We have then made the profile likelihood of each spectrum with a Log-Parabola (LP) and with a MBPWL with 2 nodes. The comparison between both results is shown in \autoref{fig:1ES_MBPWL_prof}.

Thanks to the pile-up effect when deabsorbing too much EBL (high values of $\alpha$), the upper limits obtained with the MBPWL and the LP are nearly the same, but, when removing the assumption that the intrinsic spectrum of the source is a LP, the lower limit completely disappears since the 1ES1011+496 absorbed spectra shape can be better fitted with the MBPWL than with the LP. 
\subsection{"Concave" EBL method}
The second method that we developed consists of attributing all the EBL curvature to the source, except the inflection point. In this case, when changing $\alpha$, we will not scale all the EBL, instead, we will only change the depth of the \textit{wiggle}. To do that, we defined a concave version of the EBL transmissivity curve $e-^{\tau}$ vs. E (in log-log representation) by replacing the convex part with a PWL tangent to the curve on both sides of the \textit{wiggle}. With this modified EBL and the following formula, we can modify the depth of the \textit{wiggle} by changing $\alpha$:

\begin{equation}\label{eq:tauprime}
    \frac{dF}{dE} =g(E)\cdot e^{-\tau'(E,z)} \cdot e^{-\alpha(\tau(E,z) - \tau'(E,z))}
\end{equation}

Where $g(E)$ is the fit function for the intrinsic spectra, $\alpha$ is the EBL scale, $\tau(E,z)$ is the EBL optical depth of the model and $\tau'(E,z)$ is the modified EBL optical depth that has no inflection points.

\begin{figure}
    \centering
    \includegraphics[width=\linewidth]{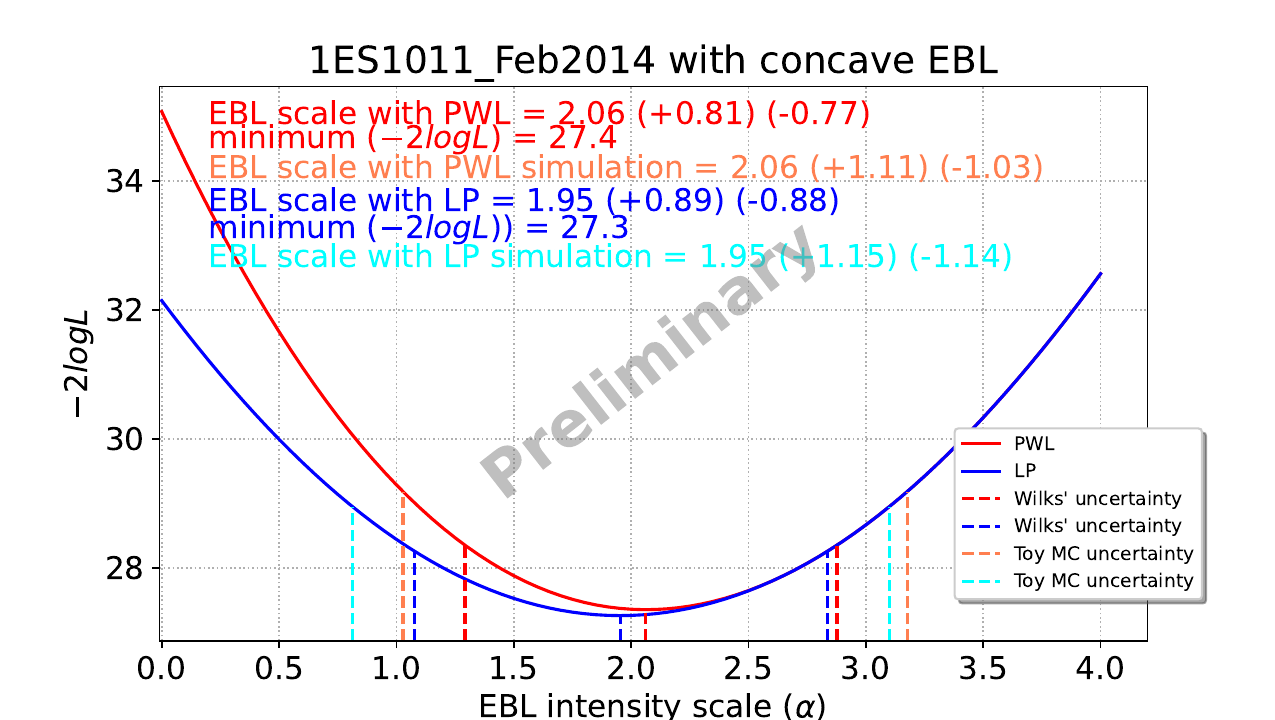}
    \caption{Profile likelihood of the 1ES1011+496 data fitted by a PWL and LP using the concave EBL method.}
    \label{fig:concave}
\vspace{-0.2cm}
\end{figure}

We have tested this method with 1ES1011+496 data and simulation and obtained the results shown in \autoref{fig:concave}.

This method does not give very constraining upper or lower limits to the EBL intensity with the current telescopes but more energy resolution and statistics, like the ones provided by the next generation of telescopes, will improve the results obtained.

\section{Conclusions}
We revised the assumptions and methods used to constrain the EBL density using gamma-ray observations.
We have made an open-source Monte Carlo simulation to run multiple realizations of the same observation. This highlighted the importance of running Monte Carlo simulations for this kind of study since the conditions needed for Wilks' theorem may not be fulfilled. It also showed how previous studies (not only the MAGIC ones) have underestimated the uncertainty of the results. This is probably due to systematics, using too simple spectral models or parameters reaching limits during the minimization. Thanks to the simulation we have been able to compute more accurate (although less restrictive) uncertainties for the real data results.\\
We have also developed two new methods to get EBL intensity constraints with fewer assumptions on the intrinsic spectra of the source. Both aim to look for the \textit{wiggle} found in the EBL transmissivity curve since it is the only feature of the EBL transmissivity that, in principle, could be attributed to the intrinsic spectra of the source. The two methods are:
\begin{itemize}
    \item The first method uses a generic concave function (MBPWL) to fit the spectra and look for the \textit{wiggle}. Results with this method completely remove the lower limit on the EBL intensity scale while maintaining the upper limits (compared with a LP fit) due to the pile-up effect that happens when deabsorbing too much EBL. 
    \item The second method uses the nominal EBL transmissivity except at the inflection point, where we do the profile likelihood changing how deep the \textit{wiggle} is.
\end{itemize}
We have tested both methods with simulations and real data. For the current generation of telescopes, neither of them is very restrictive in obtaining lower limits while the MBPWL method is still useful for obtaining upper limits. But these methods will become better and more relevant for the next generation of telescopes where we will have more statistics and more energy resolution.

\begin{acknowledgements}
The research leading to these results has received funding from the FSE under the program Ayudas predoctorales of the Ministerio de Ciencia e Innovación  PRE2020-093561
\end{acknowledgements}

\bibliographystyle{aa}
\bibliography{bibliography}

\end{document}